# WELCH's METHOD TRANSFORM OF STRANGE ATTRACTORS OF ARGON GLOW DISCHARGE PLASMAS: SYMMETRIC ZERO POINT ENERGY FLUCTUATIONS OF PARTICLES, ANTI-PARTICLES AND VIRTUAL PARTICLES

[1]M. F. Yilmaz, [2]R. Tinaztepe, [3]B.Karlik and [4]F.Yilmaz

[1]Independent Scholar, Fremont, CA,USA

[2]Department of Basic Sciences and Humanities, College of Preparation, Imam Abdulrahman Bin Faisal University, Dammam 31451, Kingdom of Saudi Arabia

[3] Neurosurgical Simulation Research and Training Centre, Department of Neurosurgery, Montreal Neurological, Institute and Hospital, McGill University, CANADA

[4] Department of Computer Sciences, New Jersey Institute of Technology, New Jersey, NY, USA

Substantial developments in pattern recognition and machine learning offer unique advancements to study the deeper structures of plasma physics. We have employed the Kernel principal component analysis(KPCA) of Argon glow discharge plasmas over time-resolved UV-Vis spectra for different pressure scans. 3D vector fields obtained by KPCA reveal the chaotic structure, strange attractors. The presence of the cross magnetic field (B=100G) and the addition of pressure increase the degree of chaos. Lower scale vector fields show that phonon sinks and traps and cool the plasma electron temperatures. The average electron and ion density profiles are provided by the 2D particle in cell simulation. Welch's transformation of the lower to higher scale vector fields illuminates the symmetric zero-point energy fluctuations of particles, antiparticles, and virtual particles. These positive and negative energies cancel each other and the combination of these energy fields form another type of fractal structure.

## I. INTRODUCTION

Studying chaos and its route with the nonlinear dynamics has brought new perspectives and stimulated the progress of science and technology in many fields. Plasmas are the chaotic state of matter, which can offer nonlinear dynamics due to their energy level and source of energy, density, pressure, etc. In general, nonlinear plasma waves and their interaction with particles regulate the chaotic distribution of the energy through a plasma medium. Furthermore, self-ordered vortex structures of electromagnetic drift waves in a bound plasma show that plasmas are suitable testbeds to study the self-organization and nonlinear dynamics depending on the applications [1 and 2].

*Corresponding author:fthyilmaz53@gmail.com*

Glow discharges plasma configurations are uncomplicated and used as the light source of fluorescent lamps and plasma display panels.On the other hand, chaotic and quasi-chaotic dynamics generated by the glow discharge plasmas are compelling phenomena and have been investigated by numerous works. Major part of the studies was realized by utilizing the fractal dimension, correlation dimension, and Lyapunov exponent analysis over the chaotic fourier power spectrum of plasmas. These works show that the discharge plasmas are transitioned to the chaos phase in the form of strange attractors. Strange attractors are mostly the consequence of nonlinear dynamics but nonchaotic attractors also do exist [3].

Another challenging phenomenon of the glow discharge experiments is the pairing state of the dust plasma generated by the ambient environments. These formations happen in the sheath area near the cathode side by the gas discharge of the particles with the size from nano to micron. Pairing states in nano to micro dimensions have been studied experimentally and theoretically extensively. However, studies of pairing in quantum dimension plasmas or so-called quantum plasmas can still be considered minor[4]. Quantum plasmas consist of degenerate particles and sub-particles such as electrons, positrons, holes, dark solitons, and quantum vortices. The charged particles mainly suit the Fermi-Dirac statistics. On the other hand, the sub-particles such as dark solitons and quantum vortices are described by the pair of equations involving the nonlinear Schroedinger and Poisson equations[5,6 and 7] Quantum plasmas are found in many astrophysical studies. The quantization of the Larmor orbits in the interior of our Earth and giant planets and brown and white dwarf stars and neutron stars are typical example of the quantum plasmas. Recently, Yilmaz et al. showed in numerous works that linear and probabilistic pattern recognition over spectral data can reveal and study the underlying physics from the low temperature and high energy density plasmas. For example, PCA analysis of spectra of K-shell Al and L-shell Mo high energy density Mo plasmas has revealed the self-organized collective V-shaped quantized pairing structures in the presence of electron beams. LDA analysis of K-shell Al plasma spectral data resulted from Langmuir turbulence structures and stated that such turbulence correlated with the social organization of the predator-prey analogy between ion and electron oscillations. Recently, B. Karli and Yilmaz et al.,2020 stated that the 3D vector fields obtained by PCA and LDA over UV spectra of gold nanoparticles revealed the strange attractors and quantum confinement of the collective plasmonic oscillations [8,9,10,11 and 12].

Furthermore, it is well known that 2 body pairing systems agree with hydrodynamics and have been applied to strongly interacting plasmas[13]. Efimov's discovery with the spectral theory of the 3-body Schrodinger operators brought a new aspect of recombination and pairing. Efimov found that 2-body interaction is weak to bind due to agitation of quantum fluctuations which tends them to diffuse. Diffusion can break the particles apart if the attraction is not strong enough. Even in the case of strong interaction particles can only remain bound together but over large distances and such interaction is said to be resonant. Efimov's suggested that a 3-body system in which the interaction of 2-particle would be mediated by the third particle and such system would sustain a long-range of attraction to bind the particles. Efimov's theory has been experimentally confirmed in 2002 for the Bose-Einstein condensation (BEC) of cesium Since the Efimov effect reveals the energy fluctuations close to absolute zero[14,15 and 16]. The Casimir effect is also another pairing mechanism that arises due to quantum fluctuations in the field. In this mechanism, virtual particles interact with the particles and antiparticles and are often described as the paired particles and these pairs annihilate in extremely short times. For those reasons, quantum hydrodynamics is the emerging field to introduce new aspects such Efimov, Casimir effects and etc., to understand the the evolution of the different types of plasmas[17 and 18].

In this work, Kernel principal component analysis (KPCA) has been applied to spectral data of Argon glow discharge plasma in the absence and the presence of the cross magnetic field (100 G). The degree of the chaos for the obtained strange attractors by the 3D vector fields has been studied using the fractal dimension, Lyapunov exponent, and Shannon entropy. Welch's transfer has been applied to understand the transformation of the strange attractor from a lower scale vector field to a higher scale. In the meantime, 2D particle in cell (PIC) modeling of the break-down oscillations, ion, and electron density profiles are also provided. The first chapter briefly describes the experiments. The second chapter studies the KPCA and the third chapter studies spectroscopic modeling and PIC modeling. Discussion of chaos using the fractal dimension, Lyapunov exponent and Shannon entropy, and zero-point energy fluctuations are given in chapter four and the conclusions are given in chapter five.

## II. EXPERIMENTS

The experiments described in this article are performed in D.C glow discharge plasma using Argon. The breakdown voltage is formed in parallel plate configuration of copper electrodes with a diameter of 2.2 cm and spacing 1 cm, housed in a cylindrical glass tube vacuum chamber of 4.8 cm in diameter and 10 cm in length, with inlet gas discharge and vacuum outlet. The discharge chamber is applied using a 100-2000 VDC power supply, and the current was 2.0 mA. During all measurements, a continuous flow of gas through the discharge chamber was maintained. The Argonne gas was used as working gas and was fed to the chamber through a controlled flow rate. The discharge chamber was pumped down by a mechanical vacuum pump to a base pressure of 0.1-1 Torr. The dependent spectra plasmas are recorded by AvaSpec-ULS3648 between 200-1100 nm.

Figure 1 a. and b. show the power signals of Argon discharge at 590 mTorr and breakdown voltages for the considered pressures in the presence and absence of the cross magnetic fields (100 G). Magnetic fields are generated using two permanent magnets. Fig.1 and b. agree that discharge receives slightly higher voltages in the presence of the magnetic fields and observed data suit on the right side of the Paschen curve of the Argon. The influence of a cross magnetic field causes a slight increase (3 to5 %) of the breakdown voltages [19]. Obtained spectra of plasmas of Argon at 590 mTorr are shown in figure 1.c and diagnostically important lines are illustrated in the table.1 with atomic transitions [20]. The spectra show intensities of most transitions are slightly higher in the presence of the magnetic field. The most intense radiation of Argon is at 750.38 nm and transition at 309.34 nm has large line broadening.

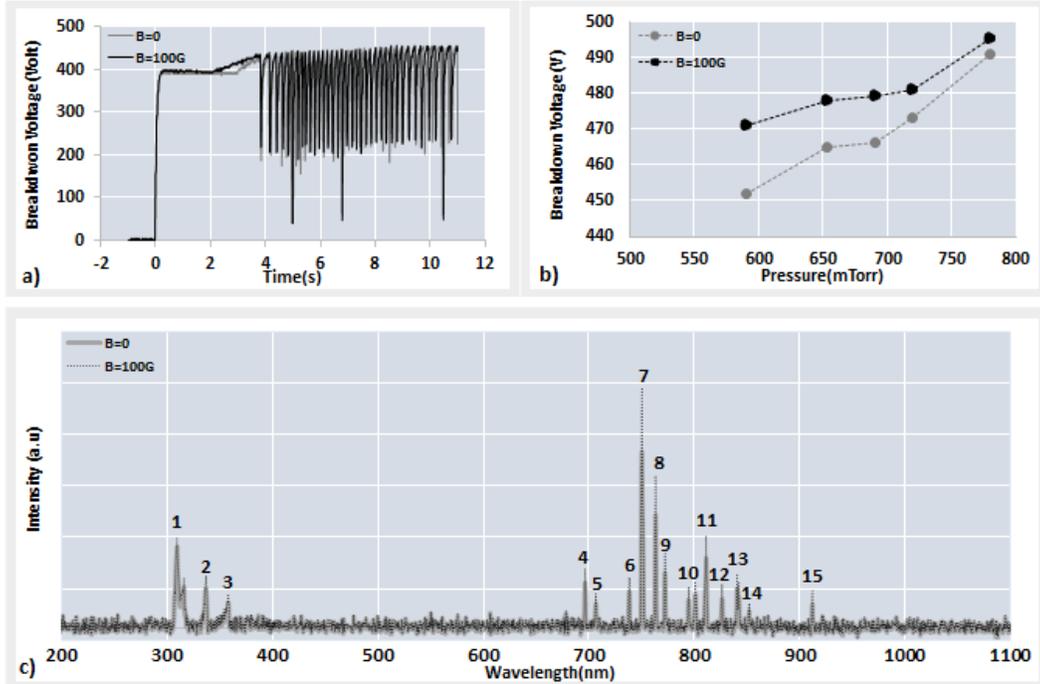

Figure 1. a) Power signal, b) Breakdown voltages and c) Spectra of Argon in the presence and absence of cross magnetic field (100G).

| Label | Ion | Transition | Wavelength(nm) |
|---|---|---|---|
| 1 | Ar II | $3s^23p^4(^3P)4d \rightarrow 3s^23p^4(^3P)4p$ | 309.34 |
| 2 | Ar II | $3s^23p^4(^1D)4d \rightarrow 3s^23p^4(^1D)4p$ | 336.55 |
| 3 | Ar II | $3s^23p^4(^3P)5d \rightarrow 3s^23p^4(^1D)4p$ | 357.07 |
| 4 | Ar I | $3s^23p^5(^2P°_{1/2})4p \rightarrow 3s^23p^5(^2P°_{3/2})4s$ | 696.54 |
| 5 | Ar I | $3s^23p^5(^2P°_{1/2})4p \rightarrow 3s^23p^5(^2P°_{3/2})4s$ | 706.72 |
| 6 | Ar I | $3s^23p^5(^2P°_{1/2})4p \rightarrow 3s^23p^5(^2P°_{3/2})4s$ | 738.39 |
| 7 | Ar II | $3s^23p^4(^1D)3d \rightarrow 3s^23p^4(^1D)4d$ | 750.38 |
| 8 | Ar I | $3s^23p^5(^2P°_{3/2})4p \rightarrow 3s^23p^5(^2P°_{3/2})4s$ | 763.51 |
| 9 | Ar I | $3s^23p^5(^2P°_{1/2})4p \rightarrow 3s^23p^5(^2P°_{1/2})4s$ | 772.42 |
| 10 | Ar I | $3s^23p^5(^2P°_{1/2})4p \rightarrow 3s^23p^5(^2P°_{1/2})4s$ | 794.81 |
| 11 | Ar I | $3s^23p^5(^2P°_{3/2})4p \rightarrow 3s^23p^5(^2P°_{3/2})4s$ | 811.53 |
| 12 | Ar I | $3s^23p^5(^2P°_{1/2})4p \rightarrow 3s^23p^5(^2P°_{1/2})4s$ | 826.45 |
| 13 | Ar I | $3s^23p^5(^2P°_{3/2})4p \rightarrow 3s^23p^5(^2P°_{3/2})4s$ | 842.46 |
| 14 | Ar I | $3s^23p^5(^2P°_{1/2})4p \rightarrow 3s^23p^5(^2P°_{1/2})4s$ | 852.14 |
| 15 | Ar I | $3s^23p^5(^2P°_{3/2})4p \rightarrow 3s^23p^5(^2P°_{3/2})4s$ | 912.29 |

Table .1 Diagnostically important transitions of Argon in the UV-Vis and NIR region.

### III.   A) SPECTROSCOPIC AND PARTICLE IN CELL MODELING

Plasma electron temperature modeling of experimental spectra is realized by the SPARTAN radiative transfer code. SPARTAN employs the nonequilibrium state of thermodynamics and assumes the electrons follow the Maxwellian velocity distributions [21]. We have simulated the charge density distribution and electric potentials of the experiment using the 2D collision-less Particle-in-Cell (PIC)method [22 and 23]. Plasma electron temperatures and the number of electrons and argon ions are inserted into the system initially to fit the breakdown potential of the experimental data. In the 2D PIC method, we have assumed all particles are moving on a rectangular region, with horizontal length 1.0 cm and vertical length 2.2 cm. The region is divided into cells by an appropriate grid, which forms intervals with side lengths: $\Delta x = \Delta y = 0.0001$ cm. The charge and mass of the particles are taken as in real. The particles' initial positions and velocities are determined by the Maxwell-Boltzmann distribution. The parameters of the Maxwell-Boltzmann distribution are determined by certain values of the particles. Below is the flow chart of the algorithm of the PIC (Figure 4).

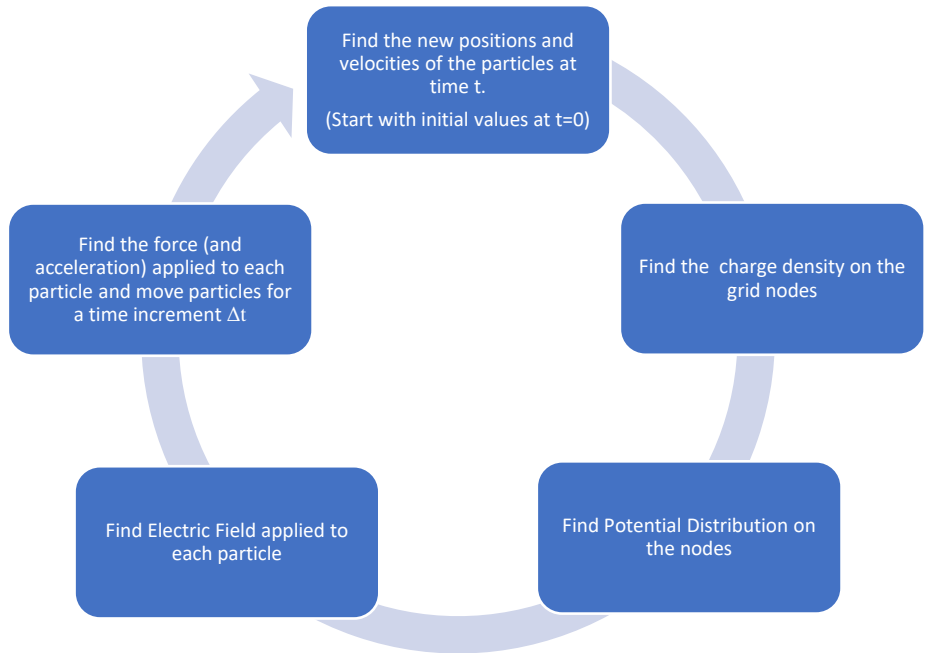

Figure 4. Flow chart of the PIC model

Fig. 5a demonstrates that the spectral modeling of Ar plasma with SPARTAN synthetic spectra gives the plasma electron temperature of 2.4 eV and 2.2 eV, in the absence and the presence of the magnetic field (100 G), respectively. PIC modeling of the electron and ion density distribution along the discharge gap is given in Figures 5-c and d. Figure 5-d also shows the pointed charge separation and ladder formation of the ion distribution in the presence of the magnetic field. The charge separation is the typical signature of the collective behavior of the plasma particles. Fig.5-e shows that the plasma electron temperature decreases in the presence of the cross magnetic field as well as the pressure increases. In the meantime, figure 5-f shows that the electron density increases in the presence of the magnetic field and it is inversely correlated by the electron temperature, which is typical characteristics of the collisionless thermal plasmas.

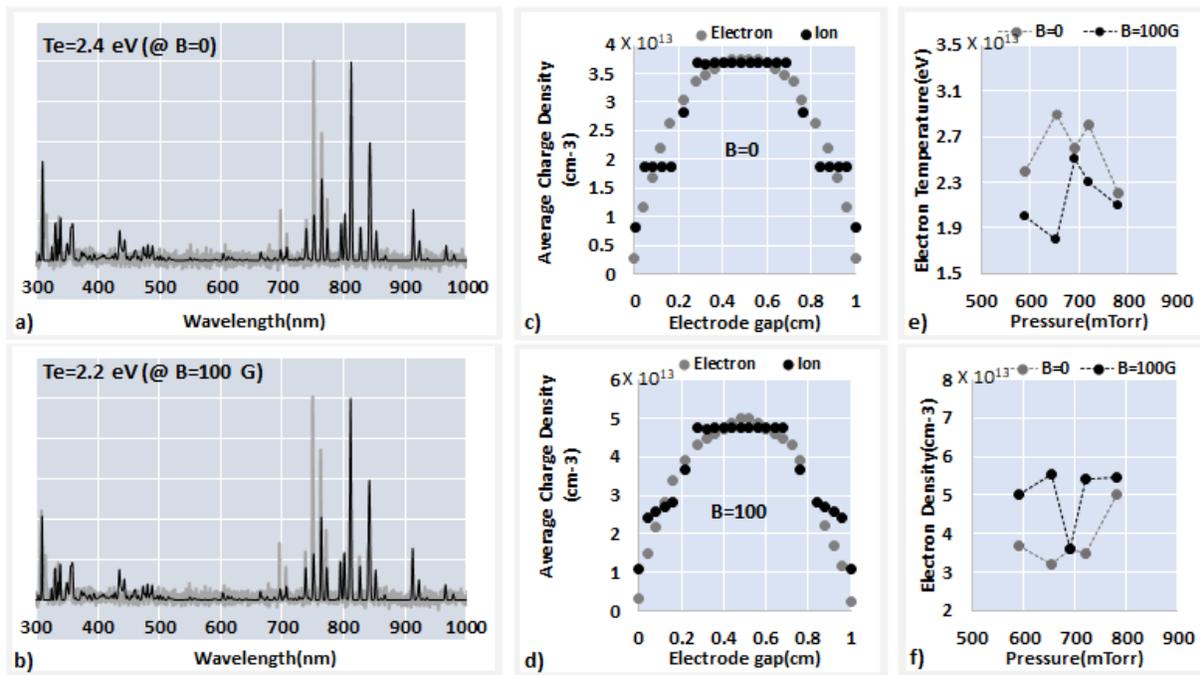

Figure 5. a and b) Spectral modeling of the electron temperatures of experiment with the 590 mTorr in the absence (B=0G) and presence of the cross magnetic field (B=100G). c and d) Electron and ion density modeling of the experiment with the 590 mTorr in the absence (B=0) and presence of the cross magnetic field (B=100G) e) Electron temperature and PSI relation in the absence (B=0) and presence of the cross magnetic field (B=100G) f) electron density and PSI relation in the absence (B=0 ) and presence of the cross magnetic field (B=100G).

## B) Kernel PCA

In general, PCA is a method that aims to keep the data set with the highest variance in high dimensional data but to provide dimension reduction while doing this. With this method, the highly correlated variables are combined to form a set of fewer artificial variables called "principal components" that make up the most variation in the data. PCA is a highly effective method for revealing the necessary information in the data. The basic logic behind PCA is to show multidimensional data with fewer variables by capturing the basic features in the data. In this study Kernelizing (or Kernel) of PCA has been used. Kernel PCA is becoming an immanent nonlinear technique applied to numerous data analysis and processing tasks.KPCA uses polynomial and Gaussian Kernel functions to projects the initial data to higher dimensional space and performs general PCA. The main advantage of Kernel PCA is that Kernel PCA suits well to the data especially with complex spatial structure. [24]

Kernel PCA leads to a latent variable technique in which maximum likelihood estimation is used to learn the model parameters and the kernel trick is still applicable. Kernel PCA is given a good reencoding of data when it lies along a non-linear manifold [25]. Consider the normalized kernel matrix $\breve{K}$ of the data (this will be of dimension N × N). The eigenvalues and eigenvectors of this matrix λj, aj are found respectively. Then the following set of features represents any data point (new or old):

$$y_j = \sum_{i=1}^{N} a_{ji} K(x, x_i), \quad j = 1, 2, \dots, N \tag{1}$$

The number of components can be limited to k < N for a more compact representation (by picking the a's corresponding to the highest eigenvalues). Here, Each $y_j$ is the coordinate of φ(x) along with one of the feature space axes $v_j$. So that,

$$v_j = \sum_{i=1}^{N} a_{ji} \varphi(x_i), \quad (sum\ goes\ to\ k\ if\ k < N) \tag{2}$$

where $v_j$ are orthogonal, the projection of φ(x) onto space spanned by them is;

$$\prod \varphi(x) = \sum_{j=1}^{N} y_j v_j = \sum_{j=1}^{N} y_j \sum_{i=1}^{N} a_{ji} \varphi(x_i), \quad (sum\ goes\ to\ k\ if\ k < N) \tag{3}$$

The reconstruction error in feature space can be evaluated as:

$$\|\varphi(x) - \prod \varphi(x)\| \tag{4}$$

In this case, the Euclidian distance in kernel space between points φ(xi) and φ(xj) is:

$$\|\varphi(x_i) - \varphi(x_j)\| = K(x_i, x_i) + K(x_j, x_j) - 2K(x_i, x_j) \tag{5}$$

In this work, Kernel PCA is applied to the data obtained for magnetic fields of B=0 and B=100 Gauss separately. For these 2 classes, 5 different pressure level with 30 spectra is considered. Hence PCA is applied to in total 5x30=150 spectra of size 2825x1 of the case B=0 and B=100 separately. Each obtained principal vector (vector spectrum)is the size of 2825x1. For each spectrum, |PC1>, |PC2>, and |PC3> vector spectra are obtained by projecting the spectra onto space spanned by these three orthogonal vectors. Hence each of 150 spectra is represented in a 3D vector space. |PC1>, |PC2> and |PC3> coordinates or so-called 3D vector fields representation of experimental spectra with the presence and absence of magnetic fields are illustrated in Figure 6. The figure shows that the vector fields reveal the pattern of strange attractors which exist in phase space [26]. The presence of the cross magnetic field rotates these attractors to form new ones.

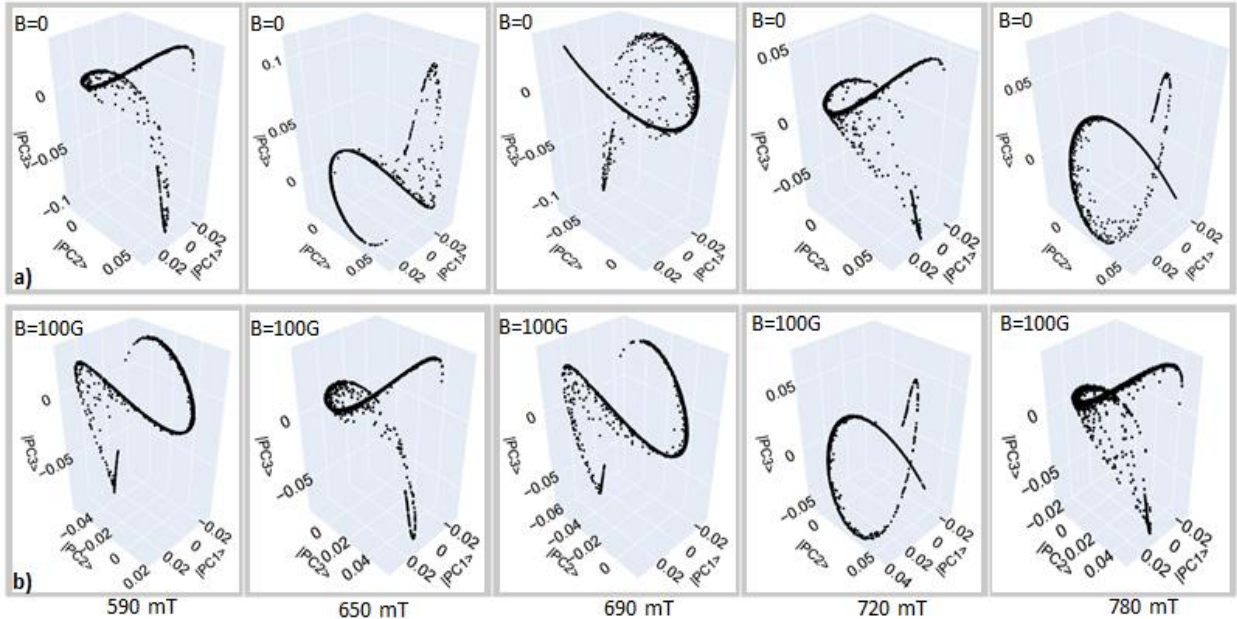

Figure 6, a and b. 3D vector fields (|PC1>, |PC2> and |PC3> coordinates) of experimental spectra in the absence and presence of the cross magnetic fields.

Figure.7 a and b show the vector spectra ( |PC1>) in which plasma waves are modulated through the radiation channel of Ar II (*309.34 nm*). According to Rabinovich et al., development of modulation instability in dissipative systems such as strange attractors due to either excitation of a large number of independent perturbations or conservative mechanisms. Zoom in regions in Figure.7 c and d around 309.34 nm shows that the oscillation is stabilized by the addition of the magnetic field. Rabinovich also states that the stabilization of the linearly tunable modes is mainly affected by the parametric energy transfer. It is well known that scattering is the main parametric mechanism in which radiation is decomposed into scattering photons and phonons [27].

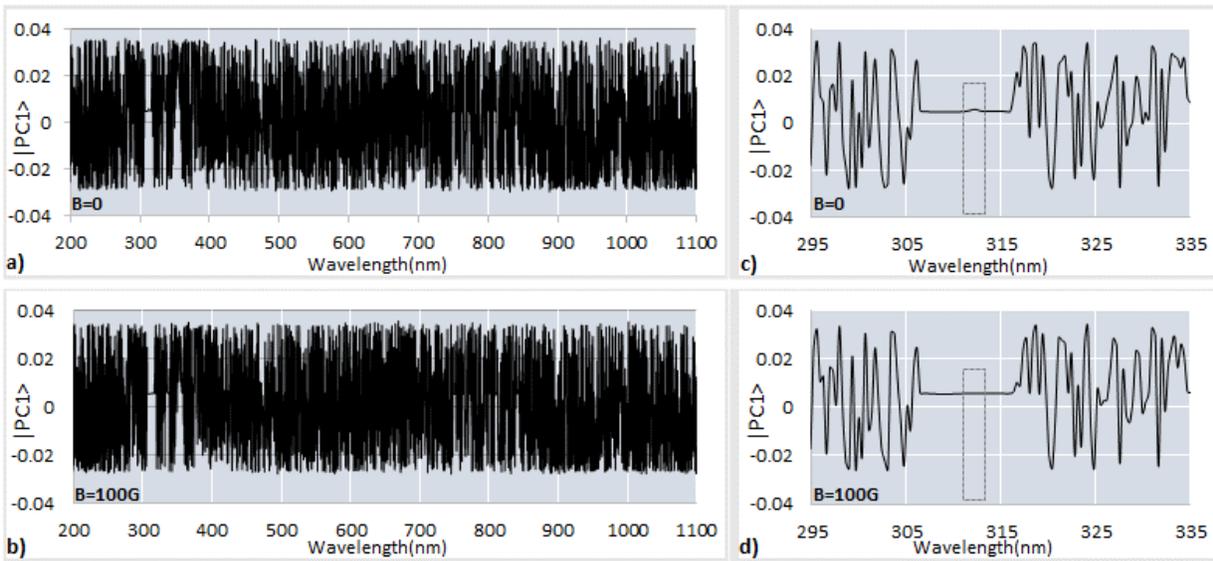

Figure 7, a and b. Vector spectra( |PC1>) in the absence and presence of the cross magnetic field. c and d) Zoom-in of the |PC1> vector spectra around Ar1 transition (309.34 nm).

## IV. DISCUSSION

Self-organization is a spontaneous occurrence of pattern formation, in which order and collective behavior arise from the interaction of sub-systems or subclasses of an initially disordered system. Such interaction is local and does not need an external agent to be controlled or so-called the interaction is self-controlled. According to Theiler, self-organization arises in dissipative systems and such systems are forbid by the law of thermodynamics and decrease their entropy. Strange attractors in phase space with chaotic trajectories are typical examples of dynamic dissipative systems. It is well known that the common element of the chaotic systems is the initial conditions of the motion and the way they set or affect the final motion[28]. The study of phenomena of strange attractors, in plasmas, has been examined to perceive the drift wave turbulence, which is very often observed in high-temperature Tokamak as well as in astrophysical plasmas[29]. Furthermore, strange attractors generated by the glow discharge plasmas also brought attention to study the chaos and routes of chaos in laboratory plasmas and interpreted for the dust plasma applications [30 and 31].

Measuring the chaos of the strange attractors is realized mostly through the fractal dimension, Lyapunov exponent, and entropy. Fractal dimension measures the complexity by the statistical index which is the ratio of the change in detail to the change in scale for the considered system or object[32]. Lyapunov exponents measure the rate of fast divergence or convergence of the infinitesimally close trajectories[33]. The thermodynamic entropy is a measure of the amount of disorder or chaos in a physical system[34]. On the other hand, Shannon information entropy is the generalized version of the classical entropy which represents the entropy as the function and identifies local or self entropies of the subsystems of the system [35]. For that reason, Shannon information entropy has received high attention to interpreting the correlation between the quantum measurements and information processing in physical systems[36].

Fractal dimension, Lyapunov exponent, and Shannon entropy of the strange attractors of the Argon glow discharge plasmas are presented in Fig. 8 a,b, and c. Figure 8.a shows that the fractal dimension increases by the pressures and however, the trend is slightly slower for the addition of a cross magnetic field in the plasma. Figures 7 c and d agree that the stabilization or damping is produced by introducing the cross magnetic field. The damping would cause the system to dissipate the energy stored in the oscillation, so the system would increase its entropy to conserve the entropy of the system at B=0. Similar trends have been observed for the Lyapunov exponent and Shannon entropy. For better understanding, the topologic images of the lower scale vector fields (|PC4>, |PC5> and |PC6>) have also been illustrated in Figure 9. Figures show that the introducing magnetic field constructs phonon sinks around 309.34 nm and 818 nm and directs plasma species towards those sinks. The trapping of the particles towards the wells has been clearly illustrated by these wells in the figure.9 b. Furthermore, Yilmaz et al., 2020 illustrated the entropy distribution of acoustic black holes generated by the interaction of laser and cold Hg plasma, in which phonon sink region recieives the maximum the entropy[37 and 38]. So, introducing a magnetic field increases the fractal dimension, Lyapunov exponent, and Shannon entropy due to arise of phonon sinks with temporal structures and these sinks cool plasma temperature down by the trapping of the plasma particles[37].

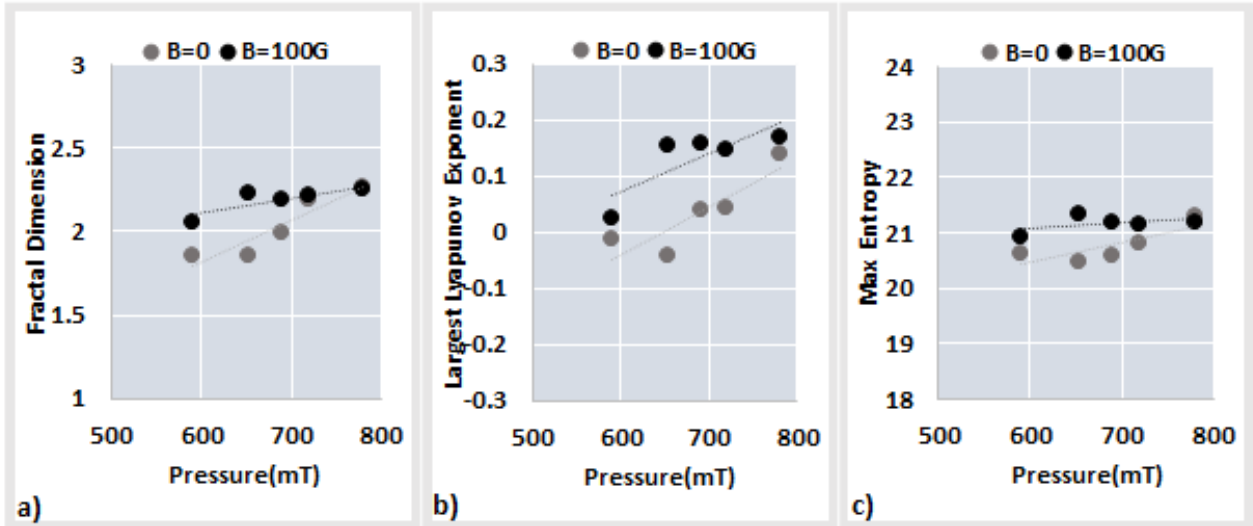

Figure 8 a) Fractal dimension, b) Lyapunov exponent and c) Maximum entropy of the strange attractors

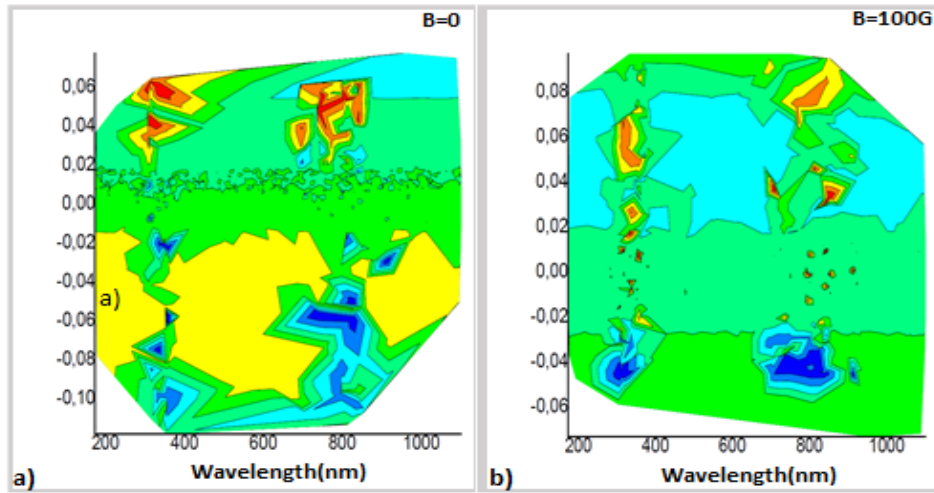

Figure 9. Topologic image of lower scale 3D vectors (|PC4>,|PC5> and |PC6>) a) in the absence b) in the presence of the cross magnetic field.

Since the quantity of the vector field is described as energy flow per unit area the $(c/4\pi)$ $(E \times H)$) or so-called Poynting vector, one can apply Welch's transfer method to evaluate energy density and the transformation of the energy density from the lower scale to higher scale vector fields. Welch's method estimates the power spectra by the concept of the periodogram spectrum. In this concept, the signal or spectra is divided into overlapping blocks by introducing channels for each block and averaging the noise of the periodogram of each block [39 and 40]. Fig.10 c and f show that Welch's transformation of the lower scale to higher scale vector fields in the absence and presence of the cross magnetic fields. Both conditions show that the transformation is based on the fluctuations of the particles and sub-particles towards the zero points in the form of symmetric triangulation. Positive energy fluctuations towards the zero-point energy are well known as the Efimov effect (pairing) of bosons, whose energy spectrum represents geometric series towards the accumulation point based on the resonant interaction of three identical bosons in 3D over the short-range of 2-body potential. The other characteristics of the Efimov pairing are that the Efimov regimes provide the short length of cutoffs in the UV regime [41]. Vector field spectra in Figures 7 c and d show the cutoffs both in the presence and absence of the magnetic fields. The negative energy density region in figures 10 c and f also reveals the antiparticles (Fermions) negative zero-point energy fluctuations, which is also known as the Dirac-sea [42]. Hybrid energy (positive-negative or negative-positive) density fluctuations point to the virtual pair states. It is well known that virtual particles tend to be in the hybrid pair states of either Virtual particle- particle (Bosons) pair of virtual particle-antiparticle (Fermions) pair states [43]. These positive and negative energies cancel each other and show that the zero-point field is the combination of all zero-point fields [44]. Besides, these fluctuations do not meet especially at zero points, which shows that zero-point energy slightly greater than minimum potential well. Such results for the observed zero-energy fields confirm that particles have must little energy to wiggle out [45 and 46].

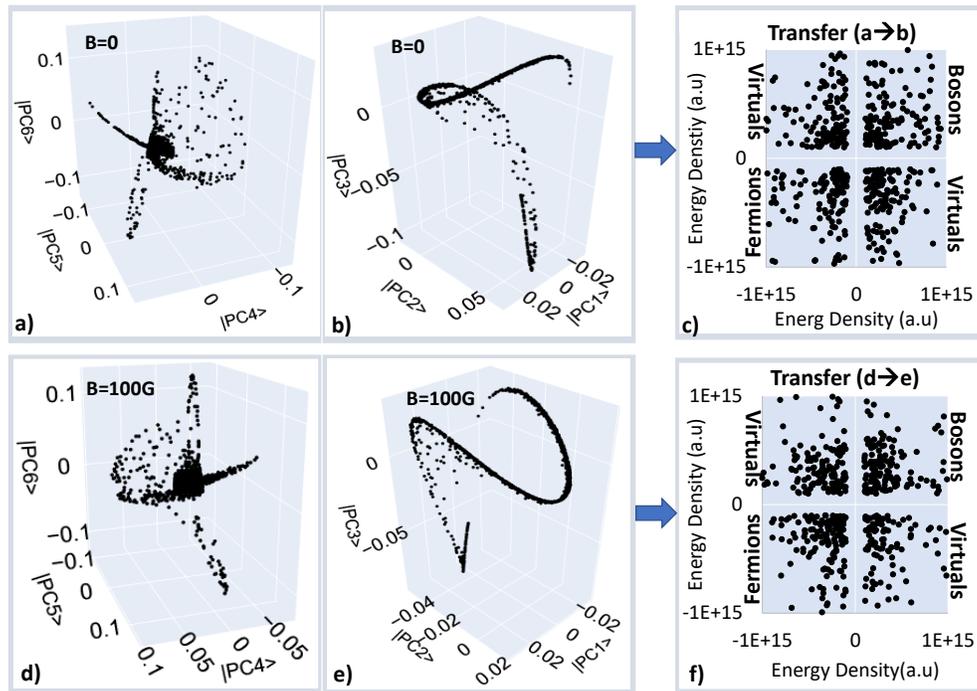

Figure 10. a) Lower scale vector field and b) Largest scale vector field in the absence of magnetic field c) Welch transformation from a to b. d) Lower scale vector field and e) Largest scale vector field in the presence of magnetic field c) Welch transformation from d to e.

## V. CONCLUSION

Our results conclude that vector fields obtained by the Kernel-PCA are alternative diagnostics to study the hidden structures of plasmas, especially the strange attractors that have spatial and temporal orientations. Such vector fields will be suitable testbeds for the reconstruction and regeneration of new vector fields through hybrid and deep machine learning techniques. Chaos study by the fractal dimension, Lyapunov exponent, and Shannon entropy follows the increasing trends with the pressure increase and the presence of the cross magnetic field. The topographic image of a lower scale vector field supports that the pressure increase and magnetic field presents a higher number of phonon sinks in the plasma which causes the increase of chaos. Phonon sinks trap the plasma species and cool the plasma. Plasma spectroscopic modeling agrees that the presence of the magnetic field decreases the plasma electron temperatures for all considered pressures. Particle in cell modeling shows the charge distribution along the discharge gap and separation which characterize the collective organization of the electrons. Welch's transformation from lower vector scale to higher vector scale shows the zero energy field fluctuations of particles, antiparticles, and hybrid states in the form of Efimov triangulation. The positive and negative energies cancel each other and show that the zero-point energy field consists of the zero-point fields of particles, anti-particles, and hybrid states. Zero-point fluctuations agree with Theiler's statement that the organization is self-controlled due to the collective interaction of the particles and virtual particles.

## ACKNOWLEDGEMENTS

This research was funded by The Scientific and Technological Research Council of Turkey with the project number of Tubitak-EEAG-113E097.